 \documentclass[10pt]{aastex63}            
 \usepackage{natbib}
\def\gapprox{\lower.4ex\hbox{$\;\buildrel >\over{\scriptstyle\sim}\;$}}
\def\lapprox{\lower.4ex\hbox{$\;\buildrel <\over{\scriptstyle\sim}\;$}}

\shortauthors{Aschwanden 2024}
\shorttitle{Galactic SOC Testing}

\begin{document}
\renewcommand{\topfraction}{0.95}
\renewcommand{\bottomfraction}{0.95}
\renewcommand{\textfraction}{0.05}
\renewcommand{\floatpagefraction}{0.95}
\renewcommand{\dbltopfraction}{0.95}
\renewcommand{\dblfloatpagefraction}{0.95}


\title{Testing the Universality of Self-Organized Criticality
in Galactic, Extra-Galactic, and Black-Hole Systems}
 
\author{Markus J. Aschwanden}
\affil{Lockheed Martin, Solar and Astrophysics Laboratory (LMSAL),
       Advanced Technology Center (ATC),
       A021S, Bldg.252, 3251 Hanover St.,
       Palo Alto, CA 94304, USA;
       e-mail: markus.josef.aschwanden@gmail.com}

\and 
\author{Ersin G\"o\v{g}\"u\c{s}} 
\affil{Sabanc\i University, Faculty of Engineering and Natural Sciences,
	Tuzla, \.Istanbul 34956, T\"urkiye;
       e-mail: ersing@sabanciuniv.edu}

\begin{abstract}
In this study we are testing whether the power law slopes 
($\alpha_F$, $\alpha_E$) of fluxes $(F)$, fluences or 
energies $(E)$ are universal in their size distributions, 
$N(F) \propto F^{-\alpha_F}$ and 
$N(E) \propto E^{-\alpha_E}$, 
in astrophysical observations of 
galactic, extragalactic, and black-hole systems. 
This is a test of fundamental importance for 
self-organized criticality (SOC) systems. 
The test decides whether (i) power laws 
are a natural consequence of the scale-freeness 
and inherent universality of SOC systems, or 
(ii) if they depend on more complex 
physical scaling laws.
The former criterion allows quantitative predictions
of the power law-like size distributions, while the later
criterion requires individual physical modeling for each
SOC variable and data set. Our statistical test, carried 
out with 61 published data sets, yields strong support 
for the former option, which implies that observed 
power laws can simply be derived from the scale-freeness 
and do not require specific physical models to 
understand their statistical distributions.
The observations show a mean and standard deviation of 
$\alpha_F=1.78\pm0.29$ for SOC fluxes, and
$\alpha_E=1.66\pm0.22$ for SOC fluences, 
and thus are consistent with the prediction
of the fractal-diffusive SOC model, with
$\alpha_F=1.80$ and $\alpha_E=1.67$.  
\end{abstract}

\section{	INTRODUCTION 				}  

This study addresses aspects of nonlinear physics and 
complexity physics applied to astrophysical phenomena.
This new field of research started with the notion of 
fractal geometry (see textbooks of Mandelbrot 1977 and 
Feder 1988). This new focus developed into the concept 
of {\sl self-organized criticality (SOC)} (Katz 1986; 
Bak et al.~1987;
1988; Bak 1996), which mostly deals with the statistics of
nonlinear events, also called avalanches or catastrophes.
On the observational side, it was found that power law
size distributions (or occurrence frequency distributions)
represent reliable hallmarks of SOC avalanche processes, 
while on the theoretical side, cellular automaton models
appear to mimic SOC processes adequately on a microscopic 
level (see textbooks by Pruessner 2012; Charbonneau 2017; 
Jensen 2023). Early applications of the SOC model to
solar physics observations and simulations were presented in
Lu and Hamilton (1991). Extensive applications to other
astrophysical phenomena were pursued thereafter 
(Aschwanden et al.~2016).

The main motivation of this endeavour is the
aim to obtain a deeper physical understanding of SOC models,
which requires theoretical models that are sufficiently
detailed to produce quantitative theoretical predictions.
At this point we observe many phenomena with power law-like
size distributions in nature (in astrophysics, magnetospheric
physics, geophysics, biophysics, sociophysics, etc.), but we 
do not understand why power laws exist, why they have specific
values for the power law slopes, which parameters have 
universality, and what is the role of waiting time 
distributions. While older SOC studies adhere to the
original Bak-Tang-Wiesenfeld (BTW) model, based on 
microscopic next-neighbor 
interactions in lattice grids, the newer SOC studies are 
quantified in terms of macroscopic physical scaling laws,
derived from the scale-free probability distribution function.
The latter model is formulated in terms of the so-called
{\sl standard fractal-diffusive self-organized criticality
model (FD-SOC)}. The previously published textbook
{\sl ``Self-Organized Criticality in Astrophysics.
The Statistics of Nonlinear Processes in the Universe''}
(Aschwanden 2011a) contains many complementary aspects
of SOC models, but does not contain any treatment of the
FD-SOC model, which has been published later on
(Aschwanden 2012a; 2014; Aschwanden et al.~2016).

The contents of this Paper are organized in observations
and data analysis of galactic, extra-galactic, and black 
hole systems (Section 2), consisting of data selection
(Section 2.1), observational instruments (Section 2.2),
observed physical phenomena (Section 2.3), a description
of the standard FD-SOC model (Section 2.4), the fluence
model (Section 2.5), alternative
SOC energy models (Section 2.6), and cellular automaton
simulations (Section 2.7). The subsequent discussion
of results (Section 3) includes the main results
of size distributions (Section 3.1), the non-universality
of waiting time distributions (Section 3.2), the effects of 
small-number statistics (Section 3.3), the choice of
inertial ranges (Section 3.4), and a comparison 
of cosmological and solar data (Section 3.5).
Conclusions are drawn in Section 4.

\section{	OBSERVATIONS AND DATA ANALYSIS 		}

\subsection{	Data Selection		}

In this study we extract size distributions 
for the flux, $N(F) \propto  F^{-\alpha_F}$, and
for the energy, $N(E) \propto  E^{-\alpha_E}$, from
a comprehensive sample of 61 published data sets, 
obtained from 
galactic, extra-galactic, and black-hole system data.
The acquisition of relevant data sets has been
accomplished by specific searches of the term
{''Self-organized criticality (SOC)'', extracted 
from the abstracts of the {\sl Astrophysics 
Data System (ADS)} (funded by NASA).
The references of these selected studies are 
tabulated in column 6 of Table 1 (for the
SOC variable of the flux), and in column 6
of Table 2 (for the SOC variable of energy or fluence),
both organized in chronological order.
The diversity of various data sets 
and the wide-spread and interdisciplinary 
scope is intended to emphasize the aspect of
SOC universality.

\subsection{	Observational Instruments 		}

Most of the observational instruments used in the selected
studies are spacecraft with detectors operating in 
soft X-ray and hard X-ray wavelengths, while a few
ground-based instruments operate in radio wavelengths.
Comparing different wavelength ranges is relevant
because all radiation processes and physical
emission mechanisms are wavelength-dependent. 
Thus, we list the covered instruments and
wavelength or energy ranges as follows.

The {\sl European X-Ray Observatory SATellite
(EXOSAT)} provided data from the {\sl medium
energy (ME)} instrument, which operates in the
energy range of 1-50 keV.   

The German-built imaging {\sl X-Ray Telescope
(XRT)} on board {\sl ROSAT} is sensitive to 
X-rays and Extreme Ultraviolet in the energy
range of (0.1-2.4 keV), providing light curves
from the {\sl High Resolution Imager (HRI)}. 

The {\sl INTErnational Gamma-Ray Astrophysics
Laboratory (INTEGRAL)} is a medium-sized ESA
mission and is optimized to surveying the
hard X-ray sky in the energy range from 15 keV to
10 MeV.

The spacecraft {\sl Chandra} with the ACIS-S/HETGS 
instrument provides 2-8 keV light curves of
Sgr A{$^*$} in the center of the Milky Way during
the {\sl Chandra X-Ray Visionary Project (XVP)}.

The {\sl Neil Gehrels Swift Observatory},
previously called {\sl Swift Gamma-Ray Burst
Explorer}, is a space observatory designed to
study {\sl Gamma-Ray Bursts (GRB)} and to monitor
the afterglow in X-rays, and UV/Visible light at
the location of a burst, covering an energy range
of 0.2-10 keV in soft X-rays, and in a wavelength
range of 170-650 nm in UV/V. {\bf The all-sky hard X-ray 
survey detector on board Swift, Burst Alet Telescope (BAT) 
detects photons in the energy range of 15-150 keV.} 

The {\sl Kepler Space Telescope} is a space telescope
launched by NASA in 2009 and lasted until 2018, 
designed to discover (occulting) Earth-sized planets 
orbiting other stars, operating in the wavelength 
range of 430-890 nm.

XMM-Newton, also known as the {\sl High Throughput
X-Ray Spectroscopy Mission} and the {\sl X-Ray
Multi-Mirror Mission} (XMM)}, is an X-ray observatory 
lauched by the {\sl European Space Emission (ESA)},
is tasked with investigating interstellar X-Ray
sources and performs joint spectroscopy in both
X-rays and visible/ultraviolet, sensitive in the
energy range of 0.1-12 keV (12-0.1 nm).

The {\sl Fermi Gamma-ray Space Telescope (FGST)}
performs all-sky gamma-ray observations and detects
cosmological phenomena such as {\sl Active Galactic
Nuclei (AGN)}, pulsars, and conducts searches for 
dark matter from a low Earth orbit, with the instrument 
{\sl Large Area Telescope (LAT)}, within an energy range of 
20 MeV-300 GeV. {\bf Fermi also contains the {\sl Gamma-Ray 
Burst Monitor (GBM)} which is sensitive to photon energies
between 8 keV and 30 MeV.} 

The {\sl Hubble Space Telescope (HST)} is one of
the most versatile telescopes in space, 
with 2.4 meter aperture, launched in 1990,  
and is still functional after 35 years. Hubble
is imaging in near-infrared ($>$750 nm), 
visible light (380-750 nm),
and ultraviolet ($<$380 nm).

The {\sl Compton Gamma Ray Observatory (CGRO)}
was a space gamma-ray observatory detecting photons with
energies from 20 keV to 30 GeV.

The {\sl Interplanetary Cometary Explorer (ICE)},
also called {\sl ISEE-3}, recorded hard X-rays at
energies of $>$30 keV.

The {\sl Rossi X-ray Timing Explorer (XTE)}
was designed to study temporal and spectral phenomena
associated with galactic and extragalactic systems, 
containing compact objects in the energy range of
2-250 keV.

Besides the previously enumerated space-based instruments, 
ground-based radio interferometers have also been used
to study SOC: 

The {\sl Australian Square Kilometre
Array Pathfinder (ASKAP)} is a radio interferometer
that is sensitive to celestial sources, such as galaxy
formation or evolution of magnetic fields in galaxies,
at wavelengths of 712-1000 MHz. 

{\sl Arecibo} radio interferometer observed
{\sl Fast Radio Bursts (FRB)} at 1.4 and 4.5 GHz.

The {\sl Very Large Array (VLA)} recorded radio data
at 3 GHz. 

The {\sl Green Bank Telescope (GBT)} is imaging
in the wavelength range of 4-8 GHz. 

Obviously, size distributions of SOC variables (e.g.,
fluxes or energies) can only be compared from different
instruments if they are tuned to identical wavelength ranges. 
However, if SOC universality holds over some 
wavelength range, the size distributions are expected
to be proportional to each other in the same wavelength
range.

\subsection{	Observed Phenomena	}

Let us summarize the nomenclature of astrophysical 
phenomena used in our study of 35 flux and 26 energy cases. 
This list (enumerated in column 5 of Tables 1 and 2),
includes about 10 different phenomena associated with
galactic, extragalactic, and black-hole systems: 

{\sl Gamma-ray bursts (GRB)} {\bf afterglow} are extremely 
energetic events occurring at cosmological distances,
exhibiting light curves with highly variable pulses,
detected at photon energies of 
{\bf $E_{phot} \lapprox 5$ keV to $\gapprox 300$ GeV.}
Prompt gamma-ray emission is believed to be generated
by internal dissipation processes from the collapse of
massive stars {\bf (for long-duration pulses)}, 
while the later afterglow is produced
through the shock wave interaction with the 
surrounding interstellar medium. Short-duration GRBs are thought
to originate from mergers of two compact objects
such as binary neutron stars or black hole-neutron
star binaries (Meszaros and Rees 1992), {\bf once
the jet has reached a sufficiently large distance
from the central engine.}.
Recent statistical SOC studies of GRBs can be found in
Wang and Dai (2013), Yi et al. (2016), Lyu et al. (2020),
Wei et al.~(2023), Peng et al.~(2023b), Li and Yang (2023), and
Maccary et al.~(2024), see also Tables 1 and 2.

{\sl X-ray flares} are common phenomena detected in the
early afterglow phase of GRBs, most prominently in the era of
the {\sl Swift} satellite. About one third of {\sl Swift} 
GRBs show X-ray flares, observed both in short and long
GRBs, typically $\Delta t \approx 10^2-10^5$ s after
the prompt emission. Theoretical interpretations 
(Yi et al.~2016) of 
X-ray flare events range from fragmentation of a 
collapsing star (King et al.~2005), fragmentation of 
an accretion disk (Perna et al.~2006), intermittent 
accretion behavior caused by a time-variable magnetic 
barrier (Proga and Zhang 2006), magnetic reconnection 
from a post-merger millisecond pulsar (Dai et al.~2006), 
to magnetic dissipation in a decelerating shell
(Giannios 2006). X-ray flares observed in Mrk 421 
might be driven by a magnetic reconnection mechanism 
(Yan et al. 2018; Giannios 2013).
Recent statistical SOC studies of galactic and
extragalactic X-ray flare phenomena are studied in
Li et al.~(2016), Wang et al.~(2017), Yang et al.~2019),
Yan et al.~(2018), Yuan et al.~(2018), Zhang et al.~(2022), 
and Wei et al.~(2023). 

{\sl Pulsar glitches:} A pulsar is a highly magnetized, 
rapidly-rotating neutron star that emits a beam of 
electromagnetic radiation. Since the beamed emission is
aligned with the magnetic axis, we observe rotationally 
modulated pulses whenever the beam axis points to the 
Earth (line-of-sight direction) during each period of 
its rapid rotation. Besides these regular periodic
pulses on time scales of milliseconds, which are measured 
with high accuracy, there occur sudden changes in rotational
frequency, aka. glitches. These anomalies are probably caused by sporadic 
unpinning of vortices that transfer momentum to the crust 
(Alpar 1977). Conservation of angular momentum produces 
then a tiny increase of the angular 
rotation rate, called {\sl ``positive spin-ups''} of the
neutron star. Recent statistical and theoretical SOC 
studies of pulsar glitches are given in
Morley and Schmidt 
(1996), Cairns (2004), Cairns et al.~(2004), 
Melatos et al.~(2008), 
Warszawski and Melatos (2008), 
Melatos et al.~(2015, 2018), 
Yu and Liu (2017), 
Kennedy et al.~(2018); and 
Gao and Wei (2024). 
{\sl Giant pulses as well as micropulses are mentioned in
studies of pulsar glitches, which may represent
two different physical mechanisms.} 

{\sl Soft gammay-ray repeaters (Magnetars): } 
Unlike GRBs, which are extra-galactic and singular events, 
a group of short duration bursts were distinct as they were 
spectrally soft and multiple bursts originated from the same 
regions of the sky. Based on these properties, {\sl Soft 
Gamma-Ray Repeaters (SGRs)} were introduced (Laros et al.~1987).
The first members of the family of SGRs were two galactic
sources (SGR 1806-20 and SGR 1900+14) and the source of the 
exceptional March 5$^{th}$ event (Mazetz et al. 1979), namely 
SGR 0526-66 in the Large Magellanic Cloud. Duncan and Thompson (1992)
proposed that a neutron star with ultrastrong magnetic field (in the 
order of $B \gapprox 10^{14}$ G) could explain extraordinatry 
characteristics of the March 5$^{th}$ event and dubbed magnetars for 
isolated neutron stars with extremely strong magnetic fields.
According to the magnetar model, repeated short duration bursts
of hard X-rays/soft gamma-rays are expected from fracturing of the
solid neutron star crust, driven by the stress of an evolving, strong 
magnetic field (Thompson and Duncan 1995). Observational confirmation
SGRs being magnetars was achieved with the discovery of the spin 
period and period derivative of SGR 1806-20 as 7.47 s and 
$7.6 \times 10^{-11}$ s s$^{-1}$, respectively (Kouveliotou et al. 1998),
therefore, yielding an inferred dipolar field strength of 
$B=8\times10^{14}$ [G]. Soon after this discovery, spin parameters
of SGR 1900+14 were measured and its magnetar nature was established
(Kouveliotou et al. 1999). There are nearly 30 magnetars identified to date.
For a compehensive review of the physics behind magnetars,
see Turolla et al.~(2015).
The recent discovery of a galactic {\sl fast radio burst (FRB)} 
associated with a hard X-ray burst from a SGR J1935+2154 has established the 
magnetar origin of at least some FRBs (Wei et al.~2021).
Recent statistical SOC studies in magnetars include
Chang et al.~(1996),
Gogus et al.~(1999, 2000, 2017),
Prieskorn and Kaaret (2012),
Wang and Dai (2013), 
Huppenkothen et al.~(2015),
Enoto et al.~(2017), 
Cheng et al.~(2020),
Wang et al.~(2021),
Wei et al.~(2021),
Zhang et al. (2023),
Peng et al.~(2023a),
Xiao et al.~(2024),  
Xie et al.~(2024), and
Gao and Wei (2024).

{\sl Blazars} are a rather extreme class of radio-loud
active galactic nuclei (AGNs), consisting of 
{\sl BL Lac objects}
(Ciprini et al.~2003), Taveccio et al. 2020a, 2020c) and
{\sl flat spectrum radio quasars} (FSQR).
Due to relativistic Doppler boosting, blazar emission
is dominated by the nonthermal emission produced by
its jet (Yan et al.~2018).
Recent statistical SOC studies on blazars, mostly 
focussing on the 3C 454.3 and Mrk 421, are described in
Zhang et al.~(2018a),
Yan et al.~(2018), and
Peng et al.~(2023a). 

An {\sl active galactic nuclei (AGN)} is a compact
region at the center of a galaxy that emits a significant
amount of energy across the electromagnetic spectrum,
with characteristics indicating that this luminosity
is not produced by the stars. The non-stellar radiation
from an AGN is believed to result from the accretion of
matter by a supermassive black hole at the center of its
host galaxy. 
Recent SOC studies on AGNs can be found in 
Lawrence and Papadakis (1993),
Leighly and O'Brien (1997), 
Xiong et al.~(2000),
Gaskell (2004), 
Uttley et al.~(2005), and
Kunjaya et al.~(2011).

{\sl X-ray Binaries} are stellar systems of a compact object 
(white dwarf, neutron star, black hole) and an evolving star,
which have been further subdivided into
{\sl low-mass X-Ray binaries} (LMXB),
{\sl high-mass X-Ray binaries} (HMXB),
{\sl supergiant fast X-ray transients} (SFXT), and
{\sl super-massive black holes} (SMBH), and
{\sl cataclysmic binaries}. 
Recent SOC studies on binaries can be found in
Uttley et al.~(2005), 
Bachev et al.~(2011),
Paizis and Sidoli (2014),
Moreira et al.~(2015),
Wang et al.~(2017),
Kennedy et al.~(2018), or
Zhang et al.~(2022).

{\sl Super Massive Black-Hole systems (SMBH)} are found 
in centers of galaxies, either in our galaxy (Sgr A$^*$), 
or other galaxies (for instance in M87; Wang et al.~2015).
Recent SOC studies on Sgr A$^*$ include
Mocanu and Grumiller (2012),
Li et al.~(2015),
Wang et al.~(2015), and
Yuan et al.~(2018).

{\sl Fast radio bursts (FRBs)} are millisecond
mysterious radio transients with anomalous high
dispersion measure (Cheng et al.~2020). FRBs
are observed mostly from cosmological distances,
which is supported by the direct localization
of FRB 121102. Events from FRBs have similar 
statistical properties as magnetar bursts. 
Recent SOC studies on FRBs can be found in
Spitler et al.~(2016),
Chatterjee et al.~(2017),
Zhang et al. (2018), 
Zhang et al. (2018b), 
Scholz et al.~(2016; 2017),
Michilli et al.~(2018),
Lu and Piro (2019),
Wang and Zhang (2019),
Gourdji et al.~(2019),
Cheng et al.~(2020),
Lin and Sang (2020),
Wei et al. (2021), 
Zhang et al. (2021),
Wang et al. (2021), and
Wang et al. (2023). 

{\sl Cosmic rays} are high-energetic particles (protons, 
helium nuclei, or electrons) that originate from within 
our Milky Way, as well as from extragalactic space, and 
are detected when they hit the Earth's atmosphere and 
produce a shower of high-energy (muon) particles. The 
energy spectrum of cosmic rays extends over a large
range of $10^9$ eV $\lapprox E \lapprox 10^{21}$ eV, with an
approximate power law slope of $\alpha_E \approx 3.0$.
A closer inspection reveals a broken power law with a ``knee''
at $E_{knee} \approx 10^{16}$ eV, which separates the cosmic 
rays accelerated inside our Milky Way (with a spectral slope of
$\alpha_{E1} \approx 2.7$) and in extragalactic space (with a
slope of $\alpha_{E2} \approx 3.3$). The sources of cosmic rays
are believed to be supernova remnants, pulsars, pulsar-wind
nebulae, active galactic nuclei, and gamma-ray burst sources.
The particles with higher energies ($E \gapprox E_{knee}$)
have a uniform and isotropic distribution over the sky and
are believed to originate mostly from active galactic nuclei).

\subsection{	The Standard FD-SOC Model	}

The fluxes $F_i, i=0, ..., n$ are basic SOC variables 
(or SOC observables), 
for which we can derive the power law slope $\alpha_F$
by using a simple theoretical model that is called
the {\sl fractal-diffusive SOC (FD-SOC)} model,
which is derived from first principles in
Aschwanden (2012a, 2014, 2015). 

There are three most general 
assumptions in the FD-SOC model: (i) the multi-fractality
in Euclidean space, (ii) classical diffusion transport, 
and (iii) incoherent emission mechanisms.
First we have to define the Euclidean space, which 
has 3 different dimensionalities, $d=1,2,3$, while 
the corresponding fractal dimensions $D_d$ are defined 
by the fractal (Hausdorff) dimension. 
The Hausdorff dimension (in 3-D space, $d=3$,)
is then defined with the fractal volume $V$,
the fractal dimension $D_V=D_3$,
and a length scale $L$ (Mandelbrot 1977, 1983, 1985),
\begin{equation}
	V = L^{D_V} \ .
\end{equation}
For sake of simplicity we deal in the following with 
the third dimension only, $d=3$, which is most relevant
in our 3-D real world. For each fractal dimension $D_V$, 
there is a range of fractal dimensions, $[D_{V,min}, 
D_{V,max}] = [d-1, d]$, which covers the range
of [2,3] for the dimensionality $d=3$.
A representative fractal
dimension is the arithmetic mean of the minimum and
maximum value,  
\begin{equation}
	D_V = {(D_{V,min}+D_{V,max}) \over 2} 
	    = d-{1 \over 2} = 2.50 \ .
\end{equation}
Since we defined a fractal dimension by a non-singular 
range, $(D_{V,min} < D_V < D_{V,max})$, our range
is consistent with the concept of multi-fractals, 
but can also be expressed by a single value, which
is consistent with mono-fractals also.

Secondly, we have to define a relationship between 
the length scale $L$ and the time scale $T$ for a
transport mechanism operating in SOC avalanches.
While the original SOC model was expressed in terms
of a cellluar automaton, driven by next-neighbor 
interactions
in the original BTW model (Bak et al.~1987),
the FD-SOC model is found to provide a suitable 
approximation that avoids the use of complex 
cellular automatons, and uses the simple physical 
scaling law of Brownian motion instead,
\begin{equation}
	T = L^{2/\beta} \ ,
\end{equation}
where $\beta=1$ corresponds to the transport process
by classical diffusion, while the propagation distance
$L$ of a SOC avalanche scales with the square-root of the 
time, $L \propto T^{1/2}$, also called Brownian motion.

Thirdly, for incoherent emission mechanisms the flux
of a SOC avalanche is proportional to the (fractal)
volume (Eq.~1),
\begin{equation}
	V \propto \left( L^{D_V} \right)^\gamma \ . 
\end{equation}
where $\gamma$ is the volume-flux coefficient,
being $\gamma=1$ for incoherent emission mechanisms, 
(e.g., bremsstrahlung, free-free emission,
gyro-resonance emission, gyro-synchrotron emission), 
and values of $\gamma \gapprox 2$ are typical for 
coherent emission mechanisms
(e.g., plasma emission in electron beams, 
loss-cone plasma emission, or {\sl Microwave 
Amplification by Stimulated Emission (MASER)}.
The incoherent emission then implies a 
proportionality of the flux $F$ to the emitting
fractal volume $V^\gamma$, i.e., 
$F \propto V^\gamma$.  

Given these 3 relationships of the FD-SOC model,
we can derive the predicted power law slopes
for SOC fluxes straightforwardly. The scale-freeness
yields the size distribution of SOC avalanche length 
scales $L$, only depending on the dimensionality 
($d=3$ in most real-world data)
\begin{equation}
	N(L) dL \propto L^{-d} dL \ .
\end{equation}
Mathematically, we perform a variable substitution 
from the length size distribution $N(L)$ to the volume  
size distribution $N(V)$, by inversion of Eq.~(4),
$L(V) = V^{1/(D_V\gamma)}$, calculating the derivative,
$(dL/dV) = V^{(1/D_V\gamma)-1}$, and inserting these two terms
into Eq.~(5) yields,
\begin{equation}
	N(V) dV = N[L(V)] {\left| {dL \over dV} \right|} dL ,
\end{equation}
\begin{equation}
	N(V) dV = L(V)^{-d/(D_V\gamma)} { V^{ 1/(D_V\gamma) - 1} } dV ,
\end{equation}
\begin{equation}
	N(V) dV = V^{-d/(D_V\gamma) + 1/(D_V\gamma) - 1 } dV
\end{equation}
\begin{equation}
	N(V) dV = V^{-[ ( d-1 ) / (D_V\gamma) ] - 1 } dV
\end{equation}
\begin{equation}
	N(V) dV = V^{-\alpha_V } dV
\end{equation}
which then yields the power law slope $\alpha_V$,
\begin{equation}
	\alpha_V=1+{(d-1) \over D_V \gamma }
		={9 \over 5} = 1.80 \ .
\end{equation}
Inserting the dimensionality $d=3$, the mean fractal
dimension $D_V=2.50$, and the incoherence coefficient $\gamma=1$,
the FD-SOC model provides a prediction of the
power law slope of fluxes $\alpha_F$ without any free parameters.
Since the fractal volume $V$ is proportional to the
flux for incoherent emission $(\gamma=1)$, it follows
that the power law slopes for fluxes and volumes,
$\alpha_F$ and $\alpha_V$, are identical,
\begin{equation}
	\alpha_F = \alpha_V \ .
\end{equation}
Consequently, the FD-SOC model predicts different slopes
for incoherent and coherent emission mechanisms, which
implies that there is no universality in the 
wavelength-dependent energy scaling laws,
while the scale-freeness predicts a universally valid
power law slope for the SOC volumes and fluxes for the
cases with incoherent emission,
i.e., $\alpha_V=\alpha_F=1.80$.  

\subsection{	Fluence Modeling		}

The detection of SOC events is generally done from
light curves of fluxes, $f(t)$, obtained by an 
automated structure detection algorithm,
or manually in form of an event catalog. Light
curves can be observed in a large number of
wavelengths in astrophysics, such as in 
gamma rays, hard X-rays,
soft X-rays, EUV, UV, visible light, radio, etc.
The central question of this study is whether the
power law-like size distributions, (also called 
occurrence frequency distributions), have an
identical power law slope $\alpha_{F,\lambda_i}$
at different wavelenghts $\lambda_i$,
which would fulfill the criterion of universality,
\begin{equation}
	 \alpha_{F\lambda_1} = \alpha_{F\lambda_2} = 
	... , \alpha_{F\lambda_i} = const \ .
\end{equation}
If the criterion of universality is fulfilled,
this would imply two possible interpretations:
(i) the same physical emission mechanism is operational 
at different wavelenghts $\lambda_i$, or (ii) the power law
slope is not a property of the wavelength-dependent
emission mechanism, but rather is caused by the scale-free
property of the SOC statistics. We will see in
the following that the second interpretation 
is more likely than the first interpretation.

The fluence is defined as the time-integrated
flux during the duration $T$ of an observed event,
\begin{equation}
	E = \int_{t_1}^{t_1+T} f(t) \ dt \ ,
\end{equation}
where $T=t_2-t_1$ is the duration of an event,
$f(t)$ is the time series of fluxes, and $E$ is the
fluence or energy during the time interval 
$[t_1, t_2]$, {\sl so that the physical unit of a 
flux $F$ is [energy/time], and the physical unit 
of a fluence $E$ is [energy]. The source area 
needs to be included in the observed distance
when the sources are located at different distances
from the observer, such as stellar or galactic objects.
Since all used datasets are selected
from publications produced by other authors, we cannot
bring all variables into self-consistent physical
units. However, since PL slopes are scale-free in
SOC datasets, we can inter-compare the size distribution
of various SOC parameters, independent of their numerical 
value or physical unit, and this way test their
universality.} 

{\sl It should be noted that flux $F$ and fluence 
or energy $E$ have different power law slopes.
In slight variance to previous versions of
the FD-SOC model, we modify the definition
of the fluence or energy $E$ as a function
of the Euclidean dimension $d$, i.e.,
$E \propto V_{peak} \propto L^{(d)}$,
which is a function of the Euclidean dimension
$d$. The prediction for the fluence is then
$a_E=a_V=1+(d-1)/d = 5/3 \sim 1.67$.
This definition is consistent with
the fractal structure of a SOC avalanche, 
because the Euclidean volume $V$ (with dimension $d$)
is essentially the envelope to the fractal volume
(with fractal dimension $D_V$), in the
spatio-temporal integration scheme of SOC events.
Consequently, the flux scales with the fractal volume
$V$, while the fluence or energy scales with the space
and time-integrated volume $V_{max}$.} 

We perform, similarly to above, a variable substitution
from the length size distribution $N(L)$ to the volume
size distribution $N(V)$, by inversion of Eq.~(7),
$L(V) = V^{1/(d\gamma)}$, calculating the derivative,
$(dL/dV) = V^{(1/d\gamma)-1}$, and inserting these two terms
into Eq.~(5) yields,
\begin{equation}
        N(V) dV = N[L(V)] {\left| {dL \over dV} \right|} dL ,
\end{equation}
\begin{equation}
        N(V) dV = L(V)^{-d} { V^{ 1/(d\gamma) - 1} } dV ,
\end{equation}
\begin{equation}
        N(V) dV = V^{-d/(d\gamma) + 1/(d\gamma) - 1 } dV
\end{equation}
\begin{equation}
        N(V) dV = V^{-[ ( d-1 ) / (d\gamma) ] - 1 } dV
\end{equation}
\begin{equation}
        N(V) dV = V^{-\alpha_E } dV
\end{equation}
which then yields the power law slope $\alpha_E$,
\begin{equation}
        \alpha_E=1+{(d-1) \over d \gamma }
                \approx \left( 2 - {1 \over d} \right)  
		\approx \left( {5 \over 3} \right) \approx 1.67 \ .
\end{equation}
where the approximative values originate
from the {\sl Standard Fractal-Diffusive SOC 
(FD-SOC)} model by assuming dimensionality $(d=3)$,
classical diffusion ($\beta=1$), and
incoherent (random) emission ($\gamma=1$).
Application of the FD-SOC model can be found in
Wang and Dai (2013).

\subsection{	Alternative SOC Energy Models	}

\subsubsection{	Isotropic Luminosity Model	}

In astrophysical observations, the distances of the 
observed objects vary from stellar distances
to extragalactic distances, and thus need to be 
corrected to a common reference distance, in order to 
compare distance-related brightness (or intensity)
variations. {\bf Energy units quoted in this Section
are given in arbitrary units ($E_{iso}, E_{rel},
E_{syn}, E_B$)}. The uncorrected burst energy in
radio bursts is generally defined in terms of
an isotropic radiation pattern, 
\begin{equation}
	E_{iso} \propto 4 \pi D_L^2 F \Delta \nu  \ ,
\end{equation}
where $D_L$ is the luminosity distance,
$F$ is the burst fluence, and $\Delta \nu$ is the
bandwidth of the observation (e.g., Wang and Zhang 2019;
Chatterjee et al.~2017; Gourdji et al. 2019; 
Lyu et al. 2020).

\subsubsection{	Relativistic Isotropic Model	}

Extra-galactic distances can be obtained
from the cosmological redshift.
In the study of Yi et al.~(2016), the prompt emission
of {\sl gamma-ray bursts (GRB)} detected by Swift
in the 0.3-10 keV energy range has been modeled 
with the isotropically radiated energy $E_{iso}$,
\begin{equation}
	E_{rel} \propto { 4 \pi D_L^2 F \over (1+z)} \ ,
\end{equation}
where $D_L$ is the luminosity distance, $F$
is the fluence of the flare, and $z$ is the
relativistic redshift correction. Similar
applications of the relativistic distance
normalization up to $z \approx 15$ have been
reported by Zhang (2018b) for the most
sensitive telescopes, such as the 500-m
Aperture Spherical radio Telescope (FAST).
Such a model was also called 
isotropic-equivalent released energy
(Maccary et al.~2024).
General-relativistic effects are incorporated
in Xiong et al.~(2000).

\subsubsection{	Dispersion Measure Model  }

The {\sl dispersion measure (DM)} is one of the
key attributes of radio pulsars and {\sl Fast
Radio Bursts (FRB)}, which is an approximate
measure of the column density of electrons between
the observer and source. The observed large spread
of fluences $E$ has motivated SOC modelers to
consider the power law volumetric rate of FRB events
per unit (isotropic) energy (Lu and Piro 2019). 
The differential size distribution is defined as,
\begin{equation}
	{dN \over dE} = A E^{-\alpha_E} 
	\qquad E \le E_{max} \ ,
\end{equation}
with a maximum energy of $E_{max}$ above which there
no FRBs occur. Combining the total number of events
$N_{tot}$ with the dispersion measure $DM_{max}$ 
and the threshold flux $F_{th}$, leads to the
estimate of the volumetric rate of FRBs 
(Lu and Piro 2019). 
The cumulative size disribution function is defined as,
\begin{equation}
	N(>E) \approx {A E^{1-\alpha_E} \over 
		\alpha_E - 1} \ .
\end{equation}
Further Bayesian analysis of the full parameter
space yields a Schechter-like model (Lu and Piro 2019),
which consists of a power law part with an exponential
drop-off,
\begin{equation}
	{dN \over dE} = {\Phi_0 \over E_{max}}
	(1 + z)^{\gamma_{Lorentz}}
	\left( {E \over E_{max}} \right)^{-\alpha_E}
	\exp{ \left( { -E \over E_{max}} \right)} \ ,
\end{equation}
where $\Phi_0$ is the volumetric rate normalization,
at a redshift $z=0$, while $(1+z)^{\gamma_{Lorentz}}$ 
is the relativistic correction. These distance 
corrections are most important for galactic and
extragalactic phenomena, such as pulsars or FRBs.

\subsubsection{	Synchrotron Model	}

The variability of synchrotron emission can be used
to constrain the magnetic field strength in the
emission region by the X-ray variability time scale
(e.g., Tavecchio et al.~1998; Yan et al.~(2018),
\begin{equation}
	t_{\rm cool} = {6 \pi m_e c \over 
	\sigma_T \gamma_{\rm Lorentz} B^2} \ ,
\end{equation}
where $t_{\rm cool}$ is the cooling time,
$B$ is magnetic field in the comoving frame,
$m_e$ is the electron rest mass,
$\sigma_T$ is the cross section of Thomson 
scattering, and $\gamma_{\rm Lorentz}$ is
the relativistic Lorentz factor of the electrons.
The observational synchrotron photon energy is
\begin{equation}
	E_{syn} \approx 1.5 \times 10^{-11}
	\gamma_{\rm Lorentz}
	{ B \over (1\ G)}
	{\delta_D \over (1+z)} \quad [{\rm keV}] \ ,
\end{equation}
where $\delta_D$ is the Doppler factor. 

\subsubsection{	Magnetic Reconnection Model	}

In solar and stellar flares, the magnetic reconnection 
process 
is often invoked, where the average magnetic energy
density $B^2/8\pi$ is used as a measure of the
SOC energy parameter (Shibata and Magara 2011),
\begin{equation}
	E_B = L^3 \left( {B^2 \over 8\pi} \right)
	\approx 3 \times 10^{30}
	\left( {B \over 10^2\ G} \right)^2
	\left( {L \over 2 \times 10^9 cm}\right)^3 \ 
	[{\rm erg}] \ .
\end{equation}
which is the typical energy for a solar flares 
(Yi et al.~2016; Peng et al.~2023). 
In order to enable tests of 
magnetic reconnection processes, measurements 
of the spatial scale $L$ and the magnetic field
density $B$ are necessary, which yield the
magnetic energy $E_B$. The parameters $L$ and $B$
are more difficult to measure than other SOC
variables, and thus are seldom available.
Other applications can be found in Cheng et al.~(2020).

\subsubsection{	Hydrodynamic Energy Model	}

There is a large number of physical model ideas that have
been sketched in the SOC literature, but have not yet
matured to a level that they could be applied to SOC data.
A hydrodynamic model has been proposed from 
advection-dominated accretion disks 
(Takeuchi and Mineshige 1997). 

\subsection{	Cellular Automaton Simulations	}

Cellular automaton models have been designed and
simulated for a variety of physical scenarios, 
such as for black hole accretion disks (Mineshige
et al.~1994a), for 1/f fluctuations in hard X-rays
from black hole objects (Mineshige et al.~1994b;
Takeuchi e al.~1995), 
for flickering of cataclysmic variables (Yonehara 
et al.~1997), for pulsar glitches (Warszawski
and Melatos 2008), for spherical geometries in
soft gamma repeater bursts driven by magnetic
reconnection (Nakazato 2014), for 1-D magnetized
grids in the afterglow of gamma-ray burst X-rays
(Harko et al.~2015), for magnetar variability
with Bayesian hierarchical models (Huppenkothen
et al.~2015), for gamma-ray blazars (Tavecchio
et al.~2020b), for extragalactic gamma-ray fluxes
(Lipari 2021), for avalanches of magnetic
flux ropes (Wang et al.~2022), or for soft
gamma-ray repeaters (Xiao et al.~2024).

One arbitrary choice in simulations with cellular
automatons is the propagation scheme of the employed
next-neighbor interactions. The original realization
in the Bak-Tang-Wiesenfeld (BTW) model employs 
4 next neighbors in a 2-D Euclidean space only,
while subsequent applications use 3 and 5 next neighbors
(Mineshige et al.~1994a; 
Negoro et al.~1995; Takeuchi et al.~1995;
Yonehara et al.~1997); Mineshige and Negoro 1999), 
2 next neighbors (Mineshige et al.~1994b; Harko
et al.~2015), 3 next neighbors (Nakazato 2014),
or 6 next neighbors (Wang et al.~2022). Since the
power law slope of size distributions strongly
depends on the dimensionality $(d=1,2,3)$ and 
the number of next neighbors, the arbitrary
choice of these pameters has no predictive
power.

Another incapability of size distributions derived
from cellular automaton algorithms is the choice of
the fitted or apparent size distribution, which includes
ideal power laws (Bak et al.~1987; Warszawski and 
Melatos et al.~2008; Lipari 2021; Wang et al.~2022; Xiao 
et al.~2024), power law with exponential cutoff 
(Mineshige et al.~1994a,b; Takeuchi et al.~1995; 
Mineshige and Negoro 1999; Nakazato 2014), and
power law with flattening (Tavecchio et al.~2020a, 2020b, 2020c;
Xie et al.~2024). Single and double power law models	
($\alpha_1, \alpha_2$) have been suggested
in Xie et al.~(2024), however without a physical model.

Most of the simulated cellular automatons are
consistent with the three-part structure of size
distributions, which includes (i) the flattening
part due to incomple sampling for the smallest
events near the detection threshold, (ii) the 
(ideal) power law part in the inertial range, 
and (iii) the exponential drop-off due to finite
system size effects for the largest events.

\section{	DISCUSSION OF RESULTS 		}

In this section we discuss the observed statistical
results of the size distributions (Section 3.1),
the non-universality of waiting time distributions
(Section 3.2), the uncertainties due to small-number
statistics (Section 3.3), the choice of (power law)
inertial ranges (Section 3.4), and comparisons
of cosmological with solar data (Setion 3.5).

\subsection{	Size Distribution Results	}

We identified 35 data sets that contain information 
on the size distribution $N(F)$ and power law slopes
$\alpha_F$ of peak fluxes $F$ (Table 1), drawn from
galactic, extragalactic, and black-hole systems published
in literature. 
The mean and standard deviation of these values is 
\begin{equation}
	\alpha_F = 1.78 \pm 0.29 \ .
\end{equation}
Likewise, we identified 25 data sets that contain 
information on the size distribution $N(E)$ and 
power law slopes $\alpha_E$ of peak energies $E$ 
(Table 2).  The mean and standard deviation of 
these values is,  
\begin{equation}
	\alpha_E = 1.66 \pm 0.22 \ .
\end{equation}
Within the statistical uncertainties, 
the observed power law slopes ($\alpha_F$ and
$\alpha_E$) agree with the theoretial prediction
of the FD-SOC model, i.e., $\alpha_F=1.80$ (Eq.~11)
and $\alpha_E=1.67$ (Eq.~20),
which is the main result of this study and  
this way confirms the universality of the
power law slopes $\alpha_F$ and $\alpha_E$.

The implication of
this result is that the statistics of SOC avalanches  
can be understood in terms of the standard FD-SOC model
that is solely based on the three assumptions of 
(i) multi-fractality, (ii) transport by classical
diffusion, and (iii) incoherent emission mechanisms.
We can now answer the fundamental questions posed in
the Introduction:
The FD-SOC model is able to predict
the existence of power laws (because power laws can be
fitted within the statistical uncertainties); to predict 
the specific values of the power law slopes (Eqs.~11, 20),
and to predict which parameters have universality 
(i.e., $\alpha_F=1.80$ for the peak flux, $\alpha_E=1.67$ 
for the dissipated energy.)  
In principle, the standard FD-SOC model can be
generalized for various dimensionalities ($d=1,2,3$),
non-classical diffusion ($\beta \neq 1$), and
coherent emission mechanisms ($\gamma > 1$),
while the standard values are ($d=3$, $\beta=1$, $\gamma=1$).
Ultimately, the original automaton concept of the BTW 
model is not a necessary condition for SOC models, 
because the microscopic next-neighbour interactions
can be replaced by the macroscopic scale-freeness
probability distribution function (Eq.~5), from which
the power law slopes can be directly derived.
Using the FD-SOC model (Section 2.4) predicts then
universality of the theoretically predicted power 
law slopes, without invoking detailed physical 
models for SOC processes. This transformation of
microscopic BTW models to macroscopic scale-free
size distributions can be considered as a major
paradigm shift of SOC models. 

\subsection{	Waiting Time Statistics		}

The common wisdom is that waiting time distributions
have an exponential fall-off for stationary (random)
distribution functions, while they have a power law 
size distribution for non-stationary distributions.
A non-stationary distribution requires at least one more
free parameter than a stationary distribution, which
can be defined in terms of a flare rate function.
A waiting time distribution function thus cannot
have universal validity for non-stationary probability
distribution functions, because of its dependence of
the flaring rate variability. Every waiting time
distribution function model can only make predictions
if there is a way to measure the flare rate, which
generally is not constant.

\subsection{	Small-Number Statistics		}

One of the prime criteria whether the observed
values of power law slopes are accurate strongly 
depends on the statistical size of the sample,
especially in the case of small-number statistics.
As a test we select only events with large-number
statistics, i.e., $n \ge 100$, and find 
$\alpha_F = 1.88$ and $\alpha_E = 1.68$, which
is however not significantly different from the 
statistics of all data sets.

\subsection{ 	Inertial Range			}

The flux range or energy range that is fitted with
a power law distribution function, also called
{\sl inertial range} is often chosen empirically
by eye, which can affect the fitted power law
slope considerably, especially if the inertial
range is relatively small. As a test we select
those subsets that have a large inertial range
of more 3 decades (which is a 
factor of $q_E=E_{max}/E_{min} > 10^3$.
Such a selection should be most reliable, 
granted that the extracted sample is 
statistically representative.
The inertial range is tabulated in column 3
of Tables 1 and 2. 
These 13 values with large energy ranges
have a mean value of $\alpha_E=1.69$, which follows
the trend of the FD-SOC predicted value of
$\alpha_E=1.67$.
 
\subsection{	Comparison with Solar Data	}

So far we obtained information on the size 
distributions and their power law slopes entirely 
from galactic, extragalactic, and black-hole systems.
However, if some universality of the power law slopes
is claimed, it is most useful to compare with 
alternative data, such as solar flares.
The mean and standard deviation of flux power law slopes
of solar data, as extracted from 21 cases in Table 2 
of Aschwanden et al.~(2016), is found to be, 
\begin{equation}
	\alpha_F^{solar} = 1.74 \pm 0.11 \ .
\end{equation}
Likewise, we identified 10 cases of fluence (or energy) 
power law slopes of solar data, as extracted from 10 
cases in Table 2 of Aschwanden et al.~(2016), with a 
mean and standard deviation of, 
\begin{equation}
	\alpha_E^{solar} = 1.56 \pm 0.11 \ .
\end{equation}
This comparison demonstrates that the power law slopes
of observed solar data agrees even better with the
theoretial predictions of the FD-SOC model, i.e., 
$\alpha_F=1.80$ (Eq.~11) and $\alpha_E=1.67$ (Eq.~20),
by using solar data rather than by using galactic data.
Moreover, the statistial uncertainties reduce from
$\sigma_F=0.29$ and $\sigma_E=0.22$ to 
$\sigma_F=0.11 = \sigma_E=0.11$, which represents
a significant reduction of the statistical uncertainty 
of the FD-SOC model. This statistical behavior is expected
since solar data can be more accurately measured at 
the much shorter solar distance than at cosmological 
scales (of galactic, extragalactic, and black-hole 
systems).

\section{	CONCLUSIONS	}	

The conclusions of this study can be summarized as follows:

\begin{enumerate}
\item{The {\sl fractal-diffusive self-organized criticality
(FD-SOC)} model is able to predict some fundamental
questions of SOC models, such as the existence
of power law-like size distribution functions, 
the specific values of the power law slopes, 
which SOC parameters have universality, and the 
role of waiting time distribution functions.}

\item{The observed medians of size distribution
functions are reported to be $\alpha_F=1.78\pm0.29$ 
for the SOC peak fluxes $F$, and $\alpha_E=1.66\pm0.22$ 
for the SOC energies $E$, which are consistent with
the theoretically predicted values of $\alpha_F
=9/5=1.80$ and $\alpha_E=1.67$. This statistical sample
indlucdes 35 cases for the flux and 26 cases for
the fluence or energy. In addition, we sort the observed
PL slopes also by SOC phenomena in Table 3, which
further corroborates the universality claimed here.}

\item{In order to understand SOC size distributions
we advocate a paradigm shift from the microscopic
next-neighbour interaction BTW model to macroscopic
scale-free probability distribution functions.
It appears that cellular automaton algorithms
are not a necessary condition for SOC behavior, 
but add unnecessary complexity. It can be 
simplified with Monte-Carlo-type simulations 
of suitable physical scaling laws.}

\item{The observed power law distribution functions 
can be fitted by a three-part model that includes
(i) the flattening due to incomplete samling of
small events, (ii) the initial range that can be 
fitted with a pure power law function, and (iii)
the steepening due to finite-system size effects
for the largest events. Such a generalized 
three-part size distribution function 
(Aschwanden 2021), 
\begin{equation}
 	N(x) dx = N_0 (x_0 + x)^{-\alpha_x} 
 		  \exp{(-{x \over x_e})} \ dx \ ,
\end{equation}
that can be fitted to an observed size distribution,
thus should include a threshold value $x_0$ and
a finite-system size limit $x_e$. This is a Pareto-type
distribution function, which converges to $N \mapsto N_0$
for $x \ll x_0$. Suitable fits require a minimum initial 
range of $E_{max}/E_{min} \approx 10^2-10^3$ decades, and a 
minimum-number statistics of $N_{min} \gapprox 10^2-10^3$.}

\item{Physical scaling law models of SOC processes
(e.g., fluence, isotropic luminosity model,
relativistic isotropic model, dispersion measure
model, synchrotron model, magnetic reconnection,
hydrodynamic model, see Section 2.6) require 
measurements of additional physical
parameters, but are not a necessary condition
to understand the statistics of SOC size distributions,
which is entirely defined by the scale-freeness
of the data in Euclidean space.}
\end{enumerate}

\acknowledgments
{\sl Acknowledgements:}
We acknowledge constructive and stimulating discussions 
(in alphabetical order)
with Sandra Chapman, Paul Charbonneau, Henrik Jeldtoft Jensen,
Adam Kowalski, Alexander Milovanov, Leonty Miroshnichenko, 
Jens Juul Rasmussen, Karel Schrijver, Vadim Uritsky, 
Loukas Vlahos, and Nick Watkins.
This work was partially supported by NASA contract NNX11A099G
``Self-organized criticality in solar physics'' and NASA contract
NNG04EA00C of the SDO/AIA instrument to LMSAL.

\clearpage

\def\ref#1{\par\noindent\hangindent1cm {#1}}

\section*{	References	}

\ref{Alpar, M.A. 1977, {\sl Pinning and Threading of Quantized 
	Vortices in the Pulsar Crust Superfluid}, 
	ApJ 213, 527.}
\ref{Aschwanden,M.J. 2011a, {\sl Self-Organized Criticality in 
	Astrophysics.
        The Statistics of Nonlinear Processes in the Universe},
        Springer-Praxis: New York, 416p.}
\ref{Aschwanden,M.J. 2012a,
        {\sl A statistical fractal-diffusive avalanche model of a
        slowly-driven self-organized criticality system},
        AA 539, A2 (15 p).}
\ref{Aschwanden, M.J. 2014,
        {\sl A macroscopic description of a generalized self-organized
        criticality systems. Astrophysical applications},
        ApJ 782, 54.}
\ref{Aschwanden, M.J. 2015,
        {\sl Thresholded powerlaw size distributions of 
	instabilities in astrophysics},
        ApJ 814, 19 (25pp).}
\ref{Aschwanden, M.J., Crosby, N., Dimitropoulou, M., Georgoulis, M.K.,
        Hergarten, S., McAteer, J., Milovanov, A., Mineshige, S.,
        Morales, L., Nishizuka, N., Pruessner, G., Sanchez, R.,
        Sharma, S., Strugarek, A., and Uritsky, V. 2016,
        {\sl 25 Years of Self-Organized Criticality: Solar and
        Astrophysics},
        Space Science Reviews 198, 47-166.}
\ref{Aschwanden, M.J. 2021,
	{\sl Finite system-size effects in self-organized criticality
	systems},
	ApJ 909:69, (12 pp).}
\ref{Atteia, J.L., Barat, C., Hurley, K., Niel, M., Vedrenne, G. et al.
        1987, {\sl  A second catalog of gamma-ray bursts: 1978-1980
        localizations from the interplanetary network}.
        ApJSS 64, 305-382.}
\ref{Bachev, R., Boeva, S., Stoyanov K., and Semkov, E. 2011,
	{\sl Intranight variability to the huge gamma-ray outburst
	of KS 1510-089},
	The Astronomers Telegram, No. 3479.}
\ref{Bak, P., Tang, C., and Wiesenfeld, K. 1987,
        {\sl Self-organized criticality - An explanation of 1/f noise},
        Physical Review Lett. 59/27, 381-384.}
\ref{Bak, P., Tang, C., and Wiesenfeld, K. 1988,
        {\sl Self-organized criticality},
        Physical Rev. A 38/1, 364-374.}
\ref{Bak, P. 1996,
        {\sl How nature works},
        Copernicus, Springer Verlag: New York.}
\ref{Beckmann, V., Borkowski, J., Courvoisier, T.J.L., Goetz, D., 
	Hudec, R., Hroch, F., Lund, N., Mereghetti, S., Shaw, S. E., 
	von Kienlin, A., and Wigger, C. 2003,
        {\sl Time resolved spectroscopy of GRB 030501 using INTEGRAL},
        AA 411, L327.}
\ref{Beloborodov, A.M. and Thompson, C. 2007,
	{\sl Corona of magnetars},
	ApJ 657/2, 967-993.}
\ref{Cairns, I.H. 2004,
        {\sl Properties and interpretations of giant micropulses
        and giant pulses from pulsars}, ApJ 610, 948-955.}
\ref{Cairns, I.H., Johnston, S., and Das, P. 2004,
        {\sl Intrinsic variability and field statistics for
        pulsars B1641-45 and B0950+08},
        MNRAS 353, 270.}
\ref{Chang, H.K., Chen, K., Fenimore, E.E., and Ho, C. 1996,
        {\sl Spectral studies of magnetic photon splitting in the March 5
        event and SGR 1806-20}, AIP Conf. Proc. 384, 921-925.}
\ref{Chatterjee, S., Law, C.J., Wharton, R.S., Burke-Spolaor, S.,
        Hessels, J.W.T. et al.~2017,
        {\sl A direct localization of a fast radio bursts and its host}, 
        Nature 541/7636, pp.58-61.}
\ref{Charbonneau, P. 2017,
        {\sl Natural complexity: A modeling handbook},
        Princeton Press: Princeton.}
\ref{Cheng, Y., Zhang, G.Q., and Wang, F.Y. 2020,
        {\sl Statistical properties of magnetar bursts and FRB 121102},
        MNRAS 491, 1498-1505.}
\ref{Ciprini, S., Fiorucci, M., Tosti, G., and Marchili, N. 2003,
        {\sl The optical variability of the blazar GV 0109+224. Hints of
        self-organized criticality},
        in {\sl High energy blazar astronomy}, ASP Conf. Proc. 229,
        (eds. L.O. Takalo and E. Valtaoja), ASP: San Francisco, p.265.}

\ref{Dai, Z.G., Wang, X.Y., Wu X.F., and Zhang B. 2006,
	{\sl X-flares from postmerger millisecond pulsars},
	Science 311, Issue 5764, 1127-1129.} 
\ref{Duncan, R. and Thompson, C. 1992, 
	{\sl Formation of Very Strongly Magnetized Neutron Stars: 
	Implications for Gamma-Ray Bursts},
	ApJ, 392, L9}	
\ref{Enoto, T., Shibata, S., Kitaguchi, T., Suwa, Y., Uchida, T. et al.~2017,
        {\sl Magnetar broadband X-ray spectra correlated with magnetic fields:
        Suzaku Archive of SGRs and AXP combned with NuSTAR, Swift, and RXTE},
        ApJSS 231/1, id. 8, 21pp.}
\ref{Feder, J. 1988,
        {\sl Fractals},
        Plenum Press: New York, 283 p.}
\ref{Gao, C.Y. and Wei, J.J. 2024,
	{\sl A comparative analysis of scale-invariant phenomena
	in repeating fast radio bursts and glitching pulsars},
	ApJ 968/1, id.40, 8pp}
\ref{Gaskell, C.M. 2004,
        {\sl Lognormal X-ray flux variations in an extreme narrow-line
        Seyfert 1 galaxy}, ApJ 612, L21.-L24.}
\ref{Giannios, D. 2013,
	{\sl Reconnection-driven plasmoids in blazars: fast flares
	on a slow envelope},
	MNRAS 431/1, p.355-363.}
\ref{Gogus, E., Woods, P.M., Kouveliotou, C., van Paradijs, J.,
        Briggs, M.S., Duncan, R.C., and Thompson, C. 1999,
        {\sl Statistical properties of SGR 1900+14 bursts},
        ApJ 526, L93-L96.}
\ref{Gogus, E., Woods, P.M., Kouveliotou, C., and van Paradijs, J. 2000,
        {\sl Statistical properties of SGR 1806-20 bursts},
        ApJ 532, L121-L124.}
\ref{Gogus, E., Lin, L., Roberts, O.J., Chakraborty, M., Kaneko, Y.,
        Gill, R. et al.~2017,
        {\sl Burst and outburst characteristics of magnetar 4U 0142+61},
        ApJ 835;68 (8 pp).}
\ref{Gourdji, K., Michilli, D., Spitler, L.G. et al. 2019,
	{\sl A sample of low-energy bursts from FRB 121102},
	ApJL 877, L19.}
\ref{Harko, T., Mocanu, G., and Stroia, N. 2015,
        {\sl Self-organized criticality in an one dimensional magnetized
        grid. Application to GRB X-ray afterglows},
        Astropys.Space.Science 357/1, id.84, 9pp.}
\ref{Huppenkothen, D., Brewer, B.J., Hogg, D.W., Murray, I., Frean, M. 
	et al.~2015,
        {\sl Dissecting magnetar variability with Bayesian hierarchical 
	models},
        ApJ 10/1, id. 66, 21pp.}
\ref{Jensen, H.J. 2023,
        {\sl Complexity Science. A study of emergence},
        Cambridge University Press, Cambridge.}
\ref{Katz,J. 1986,
        {\sl A model of propagating brittle failure in heterogeneous 
	media},
        JGR 91, 10412.}
\ref{Kennedy, M.R., Clark, C.J., Voisin, G., and Breton, R.P. 2018,
        {\sl Kepler K2 observationsof thetransitional millisecond
        pulsar PSR J1023+0038},
        MNRAS 477, 1120-1132.}
\ref{King, A., O'Brien, P.T., Goad, M.R., Osborne, J., Olsson, E.,
	and Page, K. 2005,
	{\sl Gamma-ray bursts: Restarting the engine},
	ApJ 630/2, pp.L113-L115.}
\ref{Klebesadel, R.W., Strong, I.B., and Olson, R.A. 1973,
        {\sl Observations of gamma-ray bursts of cosmic origin},
        ApJ 182, p.L85.}
\ref{Kouveliotou,C., Dieters, S., Strohmayer, T., van Paradijs, J.,
        Fishman, G.J., Meegan, C.A., Hurley, K., Kommers, J.,
        Smith, I., Frail, D., Muakami, T. 1998,
        {\sl An X-ray pulsar with a superstrong magnetic field 
	in the soft $\gamma$-ray repeater SGR 1806-20},
        Nature 393, 235-237.}
\ref{Kouveliotou,C., Strohmayer, T., Hurley, K., van Paradijs, J.,
        Finger, M.H., Dieters, S., Woods, P., Thomson, C., and Duncan, R.C.
        1999, {\sl Discovery of a magnetar associated with the soft gamma
        ray repeater SGR 1900+14}, ApJ 510, L115-L118.}
\ref{Kunjaya, C., Mahasena, P., Vierdayanti, K., and Herlie, S. 2011,
        {\sl Can self-organized critical accretion disks generate a
        log-normal emission variability in AGN?},
        Astrophys.Space.Science 336/2, pp.455-460.}
\ref{Laros, J.G., Fenimore, E.E., Klebesadel, R.W., Ateia, J.L.,
	Boar, M., Hurley, K., Niel,M., Vedrenne,, G., Kane, S.R.
	1987,
	{\sl A new type of repretitive behavior in a high-energy
	transient},
	ApJ 320/1, p.L111.}
\ref{Lawrence, A. and Papadakis, I. 1993,
	{\sl X-ray variability of active galactic nuclei: A universal
	power spectrum with lumninosity-dependent amplitude},
	ApJ 414, L85.}
\ref{Leighly, K.M., and O'Brien, P.T. 1997,
        {\sl Evidence for Nonlinear X-Ray Variability from the
        Broad-Line Radio Galaxy 3C 390.3},
        Astrophys.~J. 481, L15.}
\ref{Li, Y.P., Yuan, F., Yuan, Q. et al. 2015,
	{\sl Statistics of X-ray flares of Sagittarius A$^*$:
	Evidence for solar-like self-organized criticality
	phenomena},
	ApJ 810/2, article id. 19, 8 pp.}
\ref{Li, J., Torres, D.F., Rea, N., de Ona Wilhelmi, E. et al. 2016,
	{\sl Search for gamma-ray emission fromm AE Aquarii with
	seven years of Fermi-LAT observations},
	ApJ 832/1, article id.35, 6 pp.}
\ref{Li, X.J. and Yang, Y.P. 2023,
	{\sl Signatures of the self-organized criticality phenomenon
	in precursors of gamma-ray bursts},
	ApJ 955, Issue 2, id. L34, 7 pp.}
\ref{Lin, H.N. and Sang, Y. 2020,
        {\sl Scale-invariance in the repeating fast radio burst 121102},
        MNRAS 491, 2156-2161.}
\ref{Lipari, P. 2021,
        {\sl The origin of the power-law form of the extragalactic
        gamma-ray flux},
        Astroparticle Physics 125, id. 102507.}
\ref{Lu, E.T. and Hamilton, R.J. 1991,
        {\sl Avalanches and the distribution of solar flares},
        Astrophys. J. 380, L89-L92.}
\ref{Lu, W. and Piro, A.L. 2019,
        {\sl Implications from ASKAP fast radio burst statistics},
        ApJ 883/1, id.40, 8pp.}
\ref{Lyu, F., Li, Y.P., Hou, S.J., Wei, J.J., Geng, J.J., 
	and Wu X.F. 2020,
	{\sl Self-organized criticality in multi-pulse
	gamma-ray bursts},
	Frontiers Phys. 16/1, article id 14501.}
\ref{Mandelbrot, B.B. 1977,
        {\sl Fractals: form, chance, and dimension}, Translation of
        {\sl Les objects fractals}, W.H. Freeman, San Francisco.}
\ref{Mandelbrot, B.B. 1983,
        {\sl The fractal geometry of nature},
        W.H. Freeman, San Francisco.}
\ref{Mandelbrot, B.B. 1985,
        {\sl Self-affine fractals and fractal dimension},
        Physica Scripta 32, 257-260.}
\ref{Maccary, R., Guidorzi, C., Amati, L., Bassznini, L.,
	Bulla, M., et al. 2024,
	{\sl Distributions of energy, luminosity, duration and
	waiting times of gamma-ray burst pulses with known
	redshift detected by Fermi/GBM},
	ApJ 965, Issue 1, id. 92, 17pp.}
\ref{Melatos, A., Peralta, C., and Wyithe, J.S.B. 2008,
        {\sl Avalanche Dynamics of radio pulsar glitches},
        ApJ 672, 1103-1118.}
\ref{Melatos, A., Douglass, J.A., and Simula, T.P. 2015,
        {\sl Persistent gravitational radiation from glitching pulsars},
        ApJ 807:123, 12pp.}
\ref{Melatos, A., Howitt,G., and Fulgenzi,W. 2018,
        {\sl Size-waiting-time correlations in pulsar glitches},
        ApJ 863, 196.}
\ref{Meszaros, P. and Rees, M.J. 1992,
	{\sl Tidal heating and mass loss in neutron star binaries:
	Implications for gamma-ray burst models},
	ApJ 397, 570.}
\ref{Michilli, D., Seymour, A., Hessels, J.W.T. et al. 2018,
	{\sl An extreme magneto-ionic environment associated
	with the fast radio burst source FRB 121102},
	Nature 553, issue 7687, pp.182-185.}
\ref{Mineshige, S., Takeuchi, M., and Nishimori, H. 1994a,
        {\sl Is a black hole accretion disk in a self-organized 
	critical state?}, ApJ 435, L125-L128.}
\ref{Mineshige, S., Ouchi, B., and and Nishimori, H. 1994b,
        {\sl On the generation of 1/f fluctuations in X-rays from
        black-hole objects}, PASJ 46, 97-105.}
\ref{Mineshige, S. and Negoro, H. 1999,
        {\sl Accretion disks in the context of self-organized criticality:
        How to produce 1/f fluctuations ?},
        in {\sl High energy processes in accreting black holes},
        ASP Conf. Ser. 161, 113-128.}
\ref{Mocanu, G. and Grumiller, D. 2012,
        {\sl Self-organized criticality in boson clouds around black holes},
        Phys.Rev. D 85, Issue 10, is.105022.}
\ref{Moreira, C.A., Schneider, A.M., de Aguiar Marcus, A.M. 2015,
	{\sl Binary dynamics on star networks under external perturbations},
	Phys.Rev.E Vol. 92/4, id.042812.}  
\ref{Morley, P.D. and Schmidt, I. 1996,
	{\sl Platelet collapse model of pulsar glitches},
	Europhys.Lett. 33/2, pp.105-110.}
\ref{Nakazato, K. 2014,
        {\sl Self-organized criticality in a spherically closed cellular
        automaton: Modeling soft gamma repaeter bursts driven by magnetic
        reconnection},
        Phys.Rev.D 90/4, id.043010.}
\ref{Negoro, H., Kitamoto, S., Takeuchi, M., and Mineshige, S. 1995,
        {\sl Statistics of X-ray fluctuations from Cygnus X-1:
        Reservoirs in the disk?}, ApJ 452, L49-L52.}
\ref{Paizis, A. and Sidoli, L. 2014,
        {\sl Cumulative luminosity distributions of supergiant
        fast X-ray transients in hard X-rays},
        MNRAS 439/4, 3439-3452.}
\ref{Peng, F.K., Wang, F.Y. Shu, X.W et al. 2023,
        {\sl Self-organized criticality in solar GeV flares},
        MNRAS 518/3, 3959-3965.}
\ref{Peng, F.K., Wei, J.J., and Wang, H.Q. 2023a,
	{\sl Scale invariance in gamma-ray flares of the Sun and
	3C 454.3},
	ApJ 959/2, id.108, 8 pp.}
\ref{Peng, F.K., Wang, F.Y., Shy X.W., Hou, S.J. 2023b,
	{\sl self-organized criticality in solar GeV flares},
	MNRAS 518/3, pp.3959-3965.}
\ref{Perna, R., Bozzo, E., Stella, L. 2006,
	{\sl On the spin-up/spin-down transitions in accreting
	X-ray binaries},
	ApJ 639, 363-376.}
\ref{Prieskorn, Z. and Kaaret, P. 2012,
	{\sl Furst fluence distributions of soft gamma repeaters
	1806-20 and 1900+14 in the Rossi X-ray timing explorer PCA era},
	ApJ 755:1 (6pp).}
\ref{Proga, D. and Zhang, B. 2006,
	{\sl The late time evolution of gamma-ray bursts: ending
	hyperaccretion and producing flares},
	MNRAS 370/1, pp.L61-L65.}
\ref{Pruessner, G. 2012,
        {\sl Self-organised criticality. Theory, models and characterisation},
        Cambridge University Press: Cambridge.}
\ref{Scholz, P., Spitler, L.,G., Hessels, J.W.T. 2016,
	{\sl The repeating fast radio burst FRB 121102: 
	Multi-wavelength observations of additional bursts},
	ApJ 833, 177.}
\ref{Scholz, P., Bogdanov, S., Hessels, J.W.T. et al. 2017,
	{\sl Simultaneous X-ray, gamma-ray, and radio observations
	of the repeating fast radio burst FRB 121102},
	ApJ 846, 80.}
\ref{Shibata K. and Magara T. 2011,
	{\sl Solar Flares: Magnetohydrodynamic Processes}, 
	2011, LRSP, 8, 6.}
\ref{Spitler, L.G., Scholz, P., Hessels, J.W.T. et al. 2016,
	{\sl A repeating fast radio burst},
	Nature 531, 202.}
\ref{Takeuchi, M., Mineshige, S., and Negoro, H. 1995,
        {\sl X-ray fluctuations from black-hole objects and self 
	organization of accretion disks},
        Publ. Astron. Soc. Japan 47, 617-627.}
\ref{Takeuchi, M., Mineshige, S., and Negoro, H. 1997,
        {\sl X-ray fluctuations from advection-dominated accretion disks
        with a critical behavior}, ApJ 486, 160-168.}
\ref{Tavecchio, F., Maraschi, L., Ghisellini, G. 1998,
	{\sl Constraints on the physical parameters of TeV blazars},
	ApJ 509/2, pp.608-619.}
\ref{Tavecchio, F., Landoni, M., and Sironi, L. 2020a,
        {\sl Probing shock acceleration in BL Lac jets through X-ray
        polarimetry: the time-dependent view},
        MNRAS 498/1, pp.599-608.}
\ref{Tavecchio, F., Bonnoli, G., and Galanti, G. 2020b,
        {\sl On the distribution of fluxes of gamma-ray blazars:
        hints for a stochastic process?},
        MNRAS 497/1, pp.1294-1300.}
\ref{Tavecchio, F. and Sobacchi, E. 2020c,
        {\sl Anisotropic electron populations in BL Lac jets:
        Consequences for the observed mission},
        MNRAS 491/2, p.2198-2204.}
\ref{Thompson, C. and Duncan, R.C. 1995,
        {\sl The soft gamma repeaters as very strongly magnetized neutron
        stars. I. Radiative mechanism for outbursts}
        MNRAS 275, 255-300.}
\ref{Thompson, C. and Duncan, R.C. 1996,
        {\sl The soft gamma repeaters as very strongly magnetized neutron
        stars. II. Quiescent neutrino, X-ray, and Alfv\'en wave emission}
        ApJ 473, 322-342.}
\ref{Turolla, R., Zane, S., and Watts, A.L. 2015,
        {\sl Magnetars: the physics behind observations. A review},
        RPPh 78/11, id. 116901.}
\ref{Uttley, P., McHardy, I.M., and Baughan, S. 2005,
	{\sl Non-linear X-ray variability in X-ray binaries
	and active galaxies},
	MNRAS 359/1, pp.345-362.}
\ref{Wang, F.Y. and Dai, Z.G. 2013,
        {\sl Solar flare-like origin of X-ray flares in gamma-ray burst
        afterglows}, Nature Physics 9(8), 465-467.}
\ref{Wang, F.Y., Dai, Z.G., Yi, S.X., and Xi, S.Q. 2015,
        {\sl Universal behavior of X-ray flares from black hole systems},
        ApJSS 216/1, id.8, 8pp.}
\ref{Wang, J.S., Wang, F.Y., and Dai, Z.G. 2017,
        {\sl Self-organized criticality in type I X-ray bursts},
        MNRAS 471/3, 2517-2522.}
\ref{Wang, F.Y., and Zhang, G.Q. 2019,
        {\sl A universal energy distribution for FRB 121102},
        ApJ 882/2, id.108, 10pp.}
\ref{Wang, F.Y., Zhang, G.Q., and Dai, Z.G. 2021,
        {\sl Galactic and cosmological fast ratio bursts as
        scaled-up solar radio bursts},
        MNRAS 501/3, 3155-3161.}
\ref{Wang, W.B., Li, C., Tu, Z.L., Guo, J.H., Chen, P.F., and Wang, F.Y. 2022,
        {\sl Avalanches of magnetic flux rope in the state of self-organized
        criticality},
        MNRAS 512/2, 1567-1573.}
\ref{Wang, F.Y., Wu, Q., and Dai, Z.G. 2023,
	{\sl Repeating fast radio bursts reveal memory from minutes to 
	an hour},
	ApJL 949/2, id.L33, 9pp.}
\ref{Warszawski, L, and Melatos, A. 2008,
        {\sl A cellular automaton model of pulsar glitches},
        MNRAS 390/1, 175-191.}
\ref{Wei, Y., Zhang, B.T., and Murase, K. 2023, 
	{\sl Multiwavelength afterglow emission from bursts associated
	with magnetar flares and fast radio bursts},
	MNRAS 524/4, pp.6004-6015.}
\ref{Wei, J.J., Wu, X.F., Dai, Z.G., Wang, F.Y., Wang, P., Li, D.,
        Zhang, B. 2021,
        {\sl Similar scale-invariant behaviors between soft gamma-ray
        repeaters and an extreme epoch from FRB 121102},
        ApJ 920/2, id. 153, 7 pp.}
\ref{Xiao, S., Zhang, S.N., Xiong, S.L., Wang, P., Li, X.J.,
	Dong, A.J., Zhi, Q.J., and Li, D. 2024,
	{\sl The self-organized criticality behaviours of two new
	parameters inn SGR J1935+2154},
	MNRAS 528/2, pp.1388-1392.}
\ref{Xie,S.L., Yu, U.W., Xiong, S.L., Lin, L., Wang, P.,
	Zhao, Y., Wang, Y., han, W.L. 2024,
	{\sl Finding the particularity of the active episode of SGR
	J1935+2154 during which FRB 20200428 occurred:
	Implications from statistics of Fermi/GBM X-Ray bursts},
	ApJ 967/2, id.108, 9pp.}
\ref{Xiong, Y., Witta, P.J., and Bao, G. 2000,
        {\sl Models for accretion-disk fluctuations through self-organized
        criticality including relativistic effects},
        PASJ 52, L1097-L1107.}
\ref{Yan, D., Yang, S., Zhang, P., Dai, B., Wang, J. and Zhang, L.~2018,
        {\sl Statistical analysis of XMM-Newton X-ray flares of Mrk 421:
        Distributions of peak flux and flaring time duration},
        ApJ 864, 164 (16pp).}
\ref{Yang, H., Yuan, W., Yao, S., Li, Y., Zhang, J., Zhou, H.,
	Komossa, S., Liu, H.Y., Jin, C. 2018,
	{\sl SDSSJ211852.96-073227.5: a new gamma-ray flaring
	narrow-line Seyfert 1 galaxy},
	MNRAS 477/4, p.5127-5138.}
\ref{Yang, S., Yan,D., Dai, B., Zhang, P. and Zhu, Q. 2019,
        {\sl Statistical analysis of X-ray flares from the nucleus
        and HST-1 knot in the M87 jet},
        MNRAS 489/2, p.2685-2693.}
\ref{Yi, S., Xi, S.Q., Yu, H., Wang, F.Y., Mu, H.J.,
	Lu, L.Z., and Liang, E.W. 2016,
	{\sl Comprehensive study of the X-ray flares from
	gamma-ray bursts observed by Swift},
	ApJSS 224:20 (13pp).}
\ref{Yonehara, A., Mineshige, S., and Welsh, W.F. 1997,
        {\sl Cellular-automaton model for flickering of cataclysmic
        variables}, ApJ 486, 388-396.}
\ref{Yu, M. and Liu, Q.J. 2017,
	{\sl On the detection probability of neutron star glitches},
	MNRAS 468/3, p.3031-3041.}
\ref{Yuan, Q., Wang, Q.D., Liu, S., and Wu, K. 2018,
	{\sl A systematic Chandra study of Sgr A$^*$: II. X-ray flare
	statistis},
	MNRAS 473, 306-316.}
\ref{Zhang, J.M., Zhang, J., Yi, T.F., Huang, X.L., and Liang, E.W. 2018,
	{\sl Flux and spectral variation characteristics of 3C 454.3
	at the GeV band},
	Research in Astron. Astrophys. 18/4, id. 040.}
\ref{Zhang, J., Zhang, H.M., Yao, S., Guo, S.C., Lu, R.J. and Liang, E.W. 
	2018a,
	{\sl Jet radiation properties of 4C+49.22: from the core to
	large-scale knots},
	ApJ 865/2, id.100. 11 pp.}
\ref{Zhang, Y.G., Gaijar, V., Foster, G., Sienion, A., Cordes, J.,
	Law C. et al. 2018b,
	{\sl Fast radio burst 121102 pulse detection and periodicity:
	A machine learning approach},
	ApJ/2, id.149, 18 pp.}
\ref{Zhang, G.Q., Wang, P., Wu, Q., Wang, F.Y., Li, D., Dai, Z.B., Zhang, B.
        2021,
        {\sl Energy and waiting time distributions of FRB 121102 observed by
        FAST},
        ApJ 920L, 23.}
\ref{Zhang, W.L., Yi, S.X., Yang,U.P., and Qin,Y. 2022,
        {\sl Statistical Properties of X-Ray Flares from the Supergiant
        Fast X-Ray Transients},
        Research Astron. Astrophys. 22/6, id.065012, 8pp.}
\ref{Zhang, B. 2023,
	{\sl The physics of fast radio bursts},
	Rev.Modern.Phys. 95/3, id.035005.}

\clearpage


\begin{table}
\begin{center}
\caption{The power law slope of the size distributions of the flux
$\alpha_F$ (column 1), number of events $N_{ev}$ (column 2), 
power law range in units of decades (column 3), the instrument (column 4),
objects (column 5), and references (column 6) 
are tabulated from different published studies.}
\normalsize
\medskip
\begin{tabular}{|l|r|r|l|l|l|}
\hline
Power Law  & Number     & Power & Instrument & Object  & Reference\\
Slope of   & of         & law   &        &		&	   \\
Flux       & events     & range &        &		&	   \\
$\alpha_F$ & $N_{ev}$   & dec   &        &		&	   \\
\hline 
\hline
1.55$\pm$0.09 &  12 &   1 & EXOSAT      & AGN             & Lawrence \& Papadakis (1993)\\
1.70          &  12 &   1 & ROSAT/HRI   & 3C 390.3        & Leighly \& O'Brien (1997)\\
2.50$\pm$1.50 &  13 &   1 & INTEGRAL    & SFXT,HMXB       & Paizis \& Sidoli (2014)\\
1.65$\pm$0.17 &  38 &   3 & Chandra     & Sag A*          & Li et al. (2015)\\
2.40$\pm$0.60 &  68 &   2 & Chandra     & J1644+57        & Wang et al. (2015)\\
1.80$\pm$0.60 &  39 &   2 & Chandra     & Sgr A*          & Wang et al. (2015)\\
1.60$\pm$0.70 &  18 &   2 & Chandra     & M87             & Wang et al. (2015)\\
1.77$\pm$0.02 & 468 &   4 & Swift       & GRB             & Yi et al. (2016)\\
1.50$\pm$0.20 & 100 &   1 & Chandra     & LMXB            & Wang et al. (2017)\\
1.41          &1198 &   2 & Kepler K2   & pulsar          & Kennedy et al. (2018)\\
1.02$\pm$0.25 &  50 &   1 & XMM-Newton  & Mrk 421         & Yan et al. (2018)\\
1.54$\pm$0.02 &  34 &   2 & LAT/Fermi   & blazar 3C 454.3 & Zhang et al. (2018a)\\
1.60          &  20 &   1 & ASKAP       & FRB             & Lu and Piro (2019)\\
1.70$\pm$0.10 &   6 &   2 & VLA,Arecibo & FRB 121102      & Wang \& Zhang (2019)\\
1.63$\pm$0.19 &  10 &   2 & VLA, 3 GHz  & FRB 121102      & Chatterjee et al. (2017)\\ 
1.63$\pm$0.21 &  10 &   2 & Arecibo 1.4 GHz & FRB 121102  & Spliter et al. (2016)\\ 
1.72$\pm$0.02 &  10 &   2 & Arecibo 4.5 GHz & FRB 121102  & Michilli et al. (2018)\\ 
1.56$\pm$0.02 & 100 &   2 & GBT, 2 GHz      & FRB 121102  & Zhang et al. (2018)\\ 
1.67$\pm$0.07 &  14 &   2 & GBT, 1.4 GHzi   & FRB 121102  & Scholz et al. (2016,2017)\\ 
1.83$\pm$0.09 &  25 &   2 & Arecibo 4-8 GHz & FRB 121102  & Gourdji et al. (2019)\\ 
1.92$\pm$0.32 & 122 &   1 & Hubble HST-1    & AGN, SMBH       & Yang et al. (2019)\\
1.41$\pm$0.04 & 100 &   1 & GBT             & FRB121102        & Lin and Sang (2020)\\
2.09$\pm$0.18 & 400 &   2 & BATSE/CGRO      & GRB      & Lyu et al. (2020)\\
1.99$\pm$0.18 & 400 &   2 & BATSE/CGRO      & GRB      & Lyu et al. (2020)\\
1.66$\pm$0.06 &  93 &   2 & GRT 8 GHz       & FRB121102 & Wang et al. (2021)\\
1.82$\pm$0.20 & 112 &   3 & Fermi           & SGR 1935+2154 & Wang et al. (2021)\\
1.59$\pm$0.15 & 144 &   1 & XMM-Newton      & IGR J16418-4532 & Zhang et al. (2022)\\
1.54$\pm$0.28 & 144 &   1 & XMM-Newton      & IGR J16328-4726 & Zhang et al. (2022)\\
1.50$\pm$0.26 & 144 &   1 & XMM-Newton      & IGR J18450-0435 & Zhang et al. (2022)\\
2.19$\pm$0.12 & 122 &   2 & Swift           & GRB         & Li and Yang (2023)\\
2.44$\pm$0.07 & 243 &   1 & BATSE/CGRO      & GRB         & Li and Yang (2023)\\
2.05          & 236 &   2 & LAT/Fermi       & 3C 454.3    & Peng et al. (2023a)\\
1.91$\pm$0.35 &  39 &   2 & LAT/Fermi       & GRB         & Peng et al. (2023b)\\
1.88$\pm$0.10 &  39 &   2 & LAT/Fermi       & 3C 454.3    & Peng et al. (2023a)\\
1.95$\pm$0.02 & 158 &   4 & GBM/Fermi       & SGRJ1935+2154 & Xie et al. (2024)\\
1.47$\pm$0.42 & 974 &   3 & GBM/Fermi       & GRB         & Maccary et al. (2024)\\
\hline
\end{tabular}
\end{center}
\end{table}

\begin{table}
\begin{center}
\caption{The power law slope of the size distributions of the energy
$\alpha_E$ (column 1), instead of the flux $\alpha_F$. Otherwise similar
to Table 1.}
\normalsize
\medskip
\begin{tabular}{|l|r|r|l|l|l|}
\hline
Power Law  & Number     & Power & Instrument & Object  & Reference\\
Slope of   & of         & law   &        &		&	   \\
Energy     & events     & range &        &		&	   \\
$\alpha_E$ & $N_{ev}$   & dec   &        &		&	   \\
\hline
\hline
1.65$\pm$0.08 &  22 &   5 & BATSE      &        SGR 1900+14 & Gogus et al. (1999)\\ 
1.76$\pm$0.17 &  92 &   4 & BATSE      &        SGR 1900+14 & Gogus et al. (2000)\\ 
1.43$\pm$0.06 & 266 &   3 & RXTE       &        SGR 1900+14 & Gogus et al. (2000)\\ 
1.67$\pm$0.15 & 113 &   2 & ICE        &        SGR 1900+14 & Gogus et al. (2000)\\ 
1.77$\pm$0.01 &3000 &   3 & RXTE & SGR 1806-20     & Prieskorn \& Kaaret (2012)\\
1.94$\pm$0.03 &2000 &   3 & RXTE & SGR 1900+14     & Prieskorn \& Kaaret (2012)\\
2.06$\pm$0.15 &  83 &   4 & Swift       & GRB             & Wang \& Dai (2013)\\
1.07          &1198 &   2 & Kepler K2   & pulsar          & Kennedy et al. (2018)\\
1.73$\pm$0.25 &  82 &   2 & Chandra     & Sgr A*          & Yuan et al. (2018)\\
1.46$\pm$0.02 &  34 &   2 & LAT/Fermi & blazar 3C 454.3 & Zhang et al. (2018a)\\
1.70          &  20 &   1 & ASKAP       & FRB             & Lu and Piro (2019)\\
1.84$\pm$0.03 & 179 &   3 & RXTE        & SGRJI550-5418   & Cheng et al. (2020)\\
1.68$\pm$0.01 & 924 &   3 & RXTE        & SGR1806-20      & Cheng et al. (2020)\\
1.65$\pm$0.06 & 432 &   3 & RXTE        & SGR1900+14      & Cheng et al. (2020)\\
1.63$\pm$0.06 & 100 &   2 & RXTE        & FRB121102       & Cheng et al. (2020)\\
1.80$\pm$0.09 & 100 &   1 & GBT         & FRB121102       & Lin and Sang (2020)\\
1.54$\pm$0.09 & 400 &   2 & BATSE/CGRO  & GRB             & Lyu et al. (2020)\\
1.44$\pm$0.09 & 400 &   2 & BATSE/CGRO  & GRB             & Lyu et al. (2020)\\
1.86$\pm$0.02 &1652 &   3 & LAT/Fermi   & FRB 121102      & Zhang et al. (2021)\\
1.78$\pm$0.21 & 144 &   1 & XMM-Newton  & IGR J16418-4532 & Zhang et al. (2022)\\
1.22$\pm$0.21 & 144 &   1 & XMM-Newton  & IGR J16328-4726 & Zhang et al. (2022)\\
1.46$\pm$0.28 & 144 &   1 & XMM-Newton  & IGR J18450-0435 & Zhang et al. (2022)\\
1.90          & 236 &   2 & LAT/Fermi   & 3C 454.3        & Peng et al. (2023a)\\
1.69$\pm$0.19 &  39 &   3 & LAT/Fermi   & GRB             & Peng et al. (2023b)\\
1.69$\pm$0.01 & 158 &   4 & GBM/Fermi   & SGR J1935+2154  & Xie et al. (2024)\\
1.67$\pm$0.20 & 974 &   3 & GBM/Fermi   & GRB             & Maccary et al. (2024)\\
\hline
\end{tabular}
\end{center}
\end{table}

\begin{table}
\begin{center}
\caption{Astrophysical phenomena (column 1), number of flux data sets (column 2),
power law slope of flux distributions (column 3), number of fluence data sets (column 4),
and power law slope of fluence or energy (column 5) are listed.}
\normalsize
\medskip
\begin{tabular}{|l|l|l|l|l|}
\hline
Astrophysical                      & Number	& Power law     & Number    & Power law    \\
Phenomena                          & data sets  & Slope         & data sets & Slope        \\
				   & $N_F$      & $\alpha_F$    & $N_E$     & $\alpha_E$   \\
\hline
\hline
Galactic/Extragalactic Events      & 35         & 1.70$\pm$0.19 & 26        & 1.69$\pm$0.15\\
Gamma Ray Bursts (GRB)             & 7          & 1.83$\pm$0.20 & 5         & 1.59$\pm$0.22\\
Soft Gamma Ray Repeaters (SGR)     & 2          & 1.95$\pm$0.02 & 10        & 1.72$\pm$0.07\\
X-Ray Binaries (XB)                & 5          & 1.55$\pm$0.09 & 3         & 1.49$\pm$0.30\\
Active Galactic Nuclei (AGN)       & 3          & 1.52$\pm$0.22 &           &              \\
Blazars (BL)                       & 3          & 1.56$\pm$0.12 & 3         & 1.47$\pm$0.10\\
Fast Radio Bursts (FRB)            & 10         & 1.62$\pm$0.11 & 4         & 1.84$\pm$0.08\\
Black-Hole Systems (BH)            & 4          & 1.88$\pm$0.37 & 1         & 1.73$\pm$0.10\\
Pulsar Glitches (PLS)              & 1          & 1.41$\pm$0.04 & 1         & 1.07$\pm$0.03\\
\hline
Solar Flare Hard X-Rays (HXR)      & 20         & 1.76$\pm$0.09 & 9         & 1.48$\pm$0.11\\
Solar Flare Soft X-Rays (SXR)      & 11         & 1.94$\pm$0.12 & 5         & 1.72$\pm$0.27\\
Solar Flare EUV                    &            &               & 14        & 1.61$\pm$0.15\\
Solar Incoherent Radio Bursts      & 7          & 1.80$\pm$0.02 &           &              \\
Solar Coronal Mass Ejections (CME) & 5          & 1.93$\pm$0.38 & 7         & 1.94$\pm$0.22\\
Solar Energetic Particles (SEP)    & 7          & 1.24$\pm$0.12 & 12        & 1.34$\pm$0.09\\
Solar Wind (WIND)                  &            &               & 1         & 1.66$\pm$0.19\\
Stellar Flares                     & 33         & 2.09$\pm$0.25 &           &              \\
Stellar Flares KEPLER              & 49         & 1.98$\pm$0.08 &           &              \\
Magnetospheric Auroras             & 12         & 1.80$\pm$0.13 & 11        & 1.56$\pm$0.12\\
\hline
\hline
Observations Means                 & 17         & 1.74$\pm$0.23 & 12        & 1.64$\pm$0.16\\
\hline
FD-SOC Prediction                  &            & {\bf 1.80}    &           & {\bf 1.67}   \\
\hline
\end{tabular}
\end{center}
\end{table}


\end{document}